\documentclass[doublecol,figures]{epl2}

\title{Standing Spin Waves in an Antiferromagnetic Molecular \chem{Cr_6} Horseshoe}
 \shorttitle{Standing Spin Waves in a \chem{Cr_6} Horseshoe} 

\author{
  S. T. Ochsenbein\inst{1}\thanks{Present address: Department of Chemistry, University of Washington, Seattle, WA 98195-1700, USA}  \and O. Waldmann\inst{1} \and A. Sieber\inst{1} \and G. Carver\inst{1} \and
  R. Bircher\inst{1} \and H. U. G\"{u}del\inst{1} \and R. S. G. Davies\inst{2} \and G. A. Timco\inst{2} \and
  R. E. P. Winpenny\inst{2} \and H. Mutka\inst{3} \and F. Fernandez-Alonso\inst{4}
}
 \shortauthor{S. T. Ochsenbein \etal}

\institute{
  \inst{1} Department of Chemistry and Biochemistry, University of Bern, Freiestrasse 3, CH-3000 Bern 9,
Switzerland\\
  \inst{2} Department of Chemistry, The University of Manchester, Oxford Road, Manchester M13 9PL, UK\\
  \inst{3} Institut Laue-Langevin, 6 Rue Jules Horowitz, BP 156-38042, Grenoble Cedex 9, France\\
  \inst{4} ISIS Facility, CCLRC Rutherford Appleton Laboratory, Didcot OX11 0QX, UK
}


\abstract{ The antiferromagnetic molecular finite chain Cr$_6$ was studied by inelastic neutron scattering. The
observed magnetic excitations at 2.6 and 4.3 meV correspond, due to the open boundaries of a finite chain, to standing
spin waves. The determined energy spectrum revealed an essentially classical spin structure. Hence, various spin-wave
theories were investigated in order to assess their potential for describing the elementary excitations of finite spin
systems. }

\begin{document}

\maketitle


Since the discovery of quantum tunneling of the magnetization in the molecules \chem{Mn_{12}} and \chem{Fe_8}, the
interest in molecular nanomagnets has soared \cite{Mn12_Fe8}. Most of these molecules are characterized by dominant
nearest-neighbor Heisenberg exchange interactions between the magnetic metal centers within a molecule, and additional
weaker magnetic anisotropic terms (such as on-site anisotropy, dipole-dipole interactions, etc.). In these clusters,
the nature of the ground state and elementary spin excitations is hence determined by the Heisenberg interactions. For
a proper description of their magnetism it is thus of fundamental importance to understand the general question, what
spin ground states and excitations may emerge from Heisenberg interactions on finite spin lattices (where finite is to
be understood as to mean a few ten spin centers at most).

In this context, the antiferromagnetic rings and the Keplerate molecule \chem{Fe_{30}} have become prototypical
examples \cite{Taft94,Gat94,OW_CCR,Mueller01,Schnack01}. Their low-lying energy spectrum exhibits a remarkable
structure: In an energy vs total spin $S$ representation it consists of a set of rotational bands whose energies
increase as $E(S) \propto S(S+1)$ \cite{Anderson52,Bernu92,Schnack00,OW_SPINDYN}. Physically, the lowest rotational
band, the $L$ band, reflects the rotational degree of freedom of the N\'eel-like antiferromagnetic ground state; the
higher rotational bands, collectively called $E$ band, correspond to the (discrete) antiferromagnetic spin-wave
excitations. For the case of the antiferromagnetic rings, this picture of the excitations has been confirmed
experimentally in much detail \cite{OW_Cr8}, and also for \chem{Fe_{30}} evidence is strong \cite{Garlea06}.
Interestingly, the rotational band structure is found in other finite spin lattices, too, and is generic in this
sense. Its emergence is correlated to a basically classical spin structure \cite{Bernu92,OW_SPINDYN,OW_Cr8}.

So far, most of the understanding of the elementary excitations in Heisenberg spin clusters was inferred from exact
numerical diagonalization studies, which are obviously limited to systems with few spin sites. The classical spin
structure identified in the above systems, however, suggests to consider approaches such as spin-wave theories (SWTs),
which are standard in 3D systems, and have been demonstrated to be very capable also in 2D and 1D systems
\cite{Ivanov04}. As a great advantage SWTs would allow one to treat much larger spin clusters, and exploring their
applicability to molecular nanomagnets is hence of high interest, but no systematic study has emerged yet. Linear and
modified SWTs were recently applied to \chem{Mn_{12}} and \chem{Fe_{30}} \cite{Yama02,Chab04,Cepas05}, with no general
conclusions.

\begin{figure}
 \onefigure[scale=1]{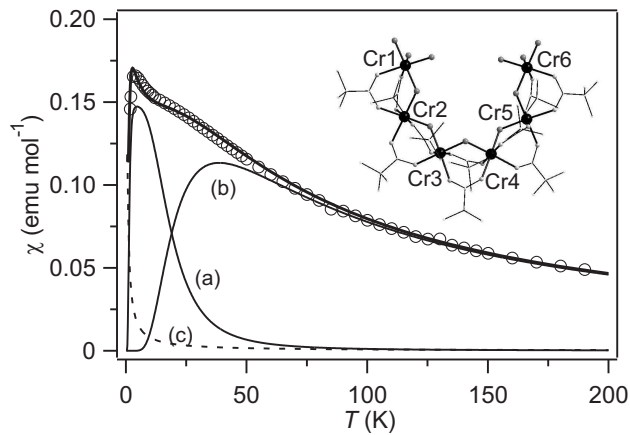}
\caption{ Magnetic susceptibility vs temperature, $\chi(T)$, of \chem{Cr_6}. Open dots: data; thick line: calculated
$\chi(T)$ as discussed in the text; thin lines: contributions from (a) the $L$-band states, (b) the remaining states,
and (c) an impurity. The inset shows the molecular structure of the \chem{[Cr_6F_{11}(O_2CCMe_3)_{10}(H_2O)]^{3-}}
anion in \chem{Cr_6}.}
 \label{fig:XvsT}
\end{figure}

The examples of the antiferromagnetic rings, \chem{Fe_{30}}, or \chem{Mn_{12}}, exhibit relatively high symmetry and
periodic boundary conditions. In this work, finite antiferromagnetic Heisenberg chains, as described by the spin
Hamiltonian
\begin{eqnarray}
  \mathcal{H} = -J \sum_{i=1}^{N-1} \vect{S}_i \cdot \vect{S}_{i+1},
  \label{eq:Ham}
\end{eqnarray}
are considered ($J < 0$ is the coupling constant, $N$ is the number of spin centers with spin length $S_i$). Finite
antiferromagnetic chains were studied extensively before, for examples see Ref.~\cite{chains}. Experimentally they were
realized by intercepting infinite chains with non-magnetic dopants. This, however, yields a distribution of chain
lengths, preventing observation of, e.g., individual spin-wave excitations \cite{chains}. The molecular compound
\chem{[^nPr_2NH_2]_3[Cr_6F_{11}(O_2CCMe_3)_{10}(H_2O)]} (or \chem{Cr_6} in short) \cite{Larsen03}, in contrast,
epitomizes a perfect finite chain: The anion of \chem{Cr_6} consists of six spin-3/2 \chem{Cr^{3+}} ions forming a
horseshoe structure, see fig.~\ref{fig:XvsT} \cite{Cr6pair}, and the generic finite-chain Hamiltonian
eq.~(\ref{eq:Ham}) is the obvious magnetic model (with $N=6$, $S_i=3/2$). All molecules in a sample are identical,
thus resolving the problem of the chain-length distribution, such that the \chem{Cr_6} compound provides unprecedented
opportunities for the experimental study of the excitations in finite antiferromagnetic chains.

The spin excitations in finite chains show similarities to rings, but also differences. In particular, due to the open
boundaries, the spin-wave excitations correspond to standing spin waves, as opposed to running waves in rings. In this
work, we used inelastic neutron scattering (INS) to study the magnetic excitations in \chem{Cr_6} experimentally, and
hence to observe the standing spin waves in a finite antiferromagnetic 1D system for the first time. We also extended
the rotational-band structure concept to finite antiferromagnetic chains, and performed various SWT calculations to
compare with the exact quantum mechanical result in order to explore the usefulness of these approaches.


A powder sample of undeuterated \chem{Cr_6} was synthesized as in \cite{Larsen03}. The magnetic susceptibility was
measured at a field of $0.1\un{T}$ with a SQUID magnetometer (Quantum Design). INS experiments in the temperature
range $1.5 - 20\un{K}$ were performed on the instruments IN5 (Institut Laue-Langevin, Grenoble, France), and IRIS
(ISIS, Didcot, UK). The data treatment employed a correction for detector efficiency by a vanadium standard. The data
shown in this work were obtained on IN5 with initial wavelengths of $\lambda = 3.0\un{{\AA}}$ and $3.8\un{{\AA}}$. The
experimental resolution at the elastic line was $0.40\un{meV}$ and $0.17\un{meV}$, respectively.

\begin{figure}
 \onefigure[scale=1]{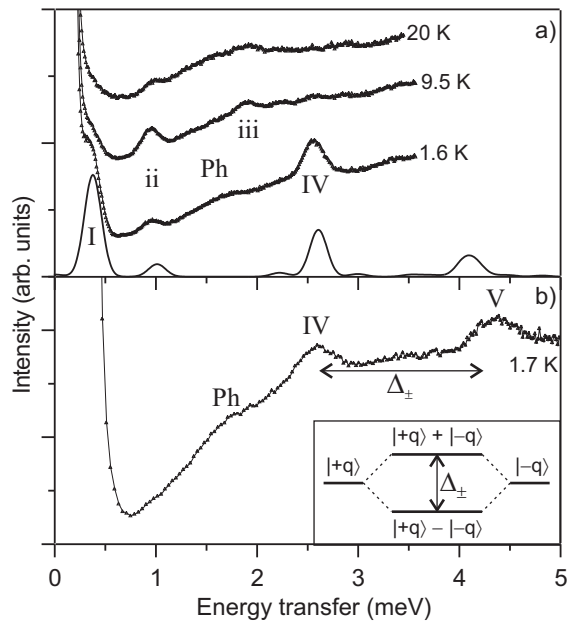}
\caption{Neutron energy-loss data of \chem{Cr_6} measured with (a) $\lambda = 3.8\un{{\AA}}$ and (b) $\lambda =
3.0\un{{\AA}}$. In (a) the curves are shifted for clarity. The solid line represents the calculated spectrum at low
temperatures ($J = -1.27\un{meV}$, $T = 1.6\un{K}$, linewidth = $0.2\un{meV}$). $\Delta_{\pm}$ denotes the $E$-band
splitting, schematically depicted in the inset.}
 \label{fig:IN5}
\end{figure}


The magnetic susceptibility vs temperature, $\chi(T)$, of \chem{Cr_6} is shown in fig.~\ref{fig:XvsT}. It exhibits a
peak at $5\un{K}$ and a shoulder at $35\un{K}$. Observation of such a behavior is unusual; a similar behavior has been
reported previously only for the antiferromagnetic wheel \chem{Cr_8Ni} \cite{Cador04}. The calculated susceptibility
(thick solid line in fig.~\ref{fig:XvsT}) was based on eq.~(\ref{eq:Ham}) with $J = -1.27\un{meV}$ as found from INS
(\textit{vide infra}), $g = 1.96$ which is typical for \chem{Cr^{3+}}, and a Curie-type contribution which accounts
for a small amount of paramagnetic impurity ($C = 0.08\un{emu~K~mol^{-1}}$). The agreement is very good, given that
$g$ and $C$ were the only adjustable parameters.

The INS data recorded on IN5 at various temperatures are shown in fig.~\ref{fig:IN5}. Six inelastic features at about
$0.4\un{meV}$ (I), $1.0\un{meV}$ (ii), $1.7\un{meV}$ (Ph), $1.9\un{meV}$ (iii), $2.6\un{meV}$ (IV), and $4.3\un{meV}$
(V) are observed. The broad band Ph and the rising background are assigned to a non-magnetic origin (incoherent
scattering of protons, phonons, internal rotations of methyl groups, etc.) \cite{phonons}. The other,
resolution-limited peaks are assigned to magnetic transitions. The temperature dependence of the $3.8\un{{\AA}}$
spectrum reveals that peaks I, IV, and V are cold transitions, while peaks ii and iii are hot transitions.

\begin{figure}
 \onefigure[scale=1]{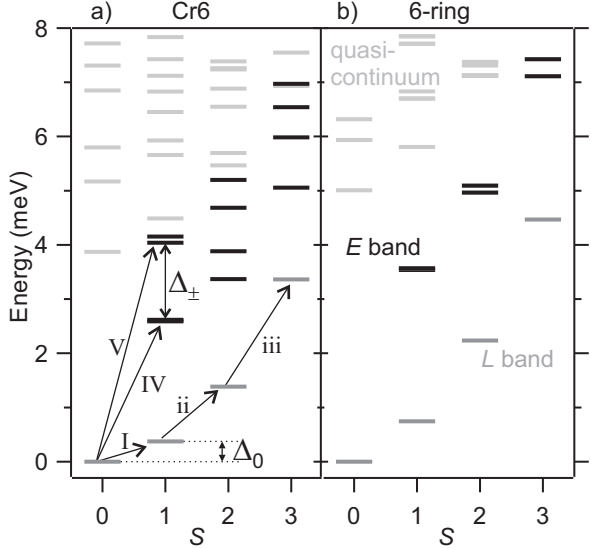}
\caption{Calculated energies vs total spin $S$ for (a) \chem{Cr_6} with $J = -1.27\un{meV}$ and (b) a $N$=6 spin-3/2
ring with $J_{ring} = \frac{5}{6} J$. The lowest states for each $S \geq 0$ form the $L$ band (dark gray levels), the 4
next-higher levels for each $S \geq 1$ the $E$ band (black levels), and the remaining states the quasi-continuum
(light gray levels).}
 \label{fig:SWvsSSW}
\end{figure}

The INS spectra were simulated by numerically diagonalizing eq.~(\ref{eq:Ham}) and calculating the intensities from the
wavefunctions using the formulas of Ref.~\cite{OW_INS}. The solid line in fig.~\ref{fig:IN5}(a) represents the
calculated spectrum for $J = -1.27\un{meV}$. The agreement with experiment is very good, considering that the model
includes only one free parameter ($J$) \cite{smallpks}.  The calculated position of peak V at $4.1\un{meV}$ is 5\% too
low, which indicates a weak variation of the exchange coupling constants along the chain \cite{Cr6params}. These
results demonstrate that the generic antiferromagnetic chain model eq.~(\ref{eq:Ham}), while it does not describe
\chem{Cr_6} perfectly, clearly provides a very good starting point for the discussion of the physical nature of the
elementary excitations in \chem{Cr_6}.

Figure~\ref{fig:SWvsSSW}(a) presents the calculated lowest energy levels vs $S$, with the observed transitions
indicated by arrows. The three cold peaks I, IV and V correspond to transitions from the $S = 0$ ground state to the
five lowest triplets (which will be identified below as the $S = 1$ states of the $L$ and $E$ bands, respectively).
Peaks IV and V each consist of two transitions, which are too close to be resolved in the experiment. Peaks ii and iii
originate from hot transitions within the $L$ band, as indicated in fig.~\ref{fig:SWvsSSW}(a).

In the following, the low-temperature triplet excitations I, IV, and V in \chem{Cr_6} will be analyzed from two points
of view. First, \chem{Cr_6} is regarded as a hexanuclear antiferromagnetic ring with a missing bond ($J_{61}$) as a
perturbation, i.e, the Hamiltonian is written as $\mathcal{H} = -J (\sum_{i=1}^{5} \vect{S}_i \cdot \vect{S}_{i+1} +
\vect{S}_6 \cdot \vect{S}_1) + \mathcal{V}$ with $\mathcal{V} = J \vect{S}_1 \cdot \vect{S}_6$. Second, various
variants of SWT will be applied.


The situation in an antiferromagnetic $N$=6 spin-3/2 ring is briefly recalled \cite{OW_SPINDYN}. The energy spectrum
for $J_{ring} = 5/6 J$ is depicted in fig.~\ref{fig:SWvsSSW}(b) (the factor 5/6 adjusts for the different number of
antiferromagnetic bonds in the chain and ring). It is characterized by a $L$ band, an $E$ band, and the
quasi-continuum. The separation of the spectrum into $L$/$E$ bands and a quasi-continuum originates from the fact,
that the INS intensity of transitions from states of the $L$ band to states of the quasi-continuum is essentially
zero, i.e., all intensity goes into transitions within the space of the $L$ and $E$ bands. Importantly, these
properties are intimately connected to a classical spin structure \cite{Bernu92,OW_SPINDYN}.

Comparing figs.~\ref{fig:SWvsSSW}(a) and \ref{fig:SWvsSSW}(b) reveals that the concept of a $L$ band, $E$ band, and
quasi-continuum is still useful for the finite chain. This is obvious for the $L$ band, though not for the $E$ band.
However, a careful inspection of the transition matrix elements shows that almost all low-temperature intensity goes
into transitions to the states displayed in dark gray and black in fig.~\ref{fig:SWvsSSW}(a). For instance,
transitions I, IV, and V account for 97.4\% of the total intensity. Thus, although the states which form the $E$ band
in \chem{Cr_6} are not easily recognized from the energy spectrum, they are clearly identified by the transition
intensities.

For a finite antiferromagnetic chain, as compared to an antiferromagnetic ring, two differences are noted. First, the
gap between the ground state and the lowest triplet $\Delta_0$ (singlet-triplet gap) is rather small. This explains
the peak and shoulder in $\chi(T)$. Figure~\ref{fig:XvsT} disentangles the contributions to $\chi(T)$ from (a) the $L$
band (which is responsible for the peak at about $5\un{K}$), (b) all other states (which produces the shoulder at
about $35\un{K}$), and (c) the impurity. As the $L$ band is governed by $\Delta_0$, the observation of a peak
separated from a shoulder directly reflects a comparatively smaller $\Delta_0$: In the antiferromagnetic rings the
$L$-band contribution (a) is shifted towards higher temperatures and hence less separated from contribution (b), such
that only a broad maximum is observed in $\chi(T)$ \cite{Taft94}. A second important difference is the splitting of
the $E$ band into two subgroups in the $N$=6 antiferromagnetic chain [from fig.~\ref{fig:SWvsSSW}(a) this is obvious
for the triplets, but less so for $S > 1$]. The splitting of the $E$ band is a direct consequence of the different
boundary conditions for the spin waves in chains and rings.

Considering a finite antiferromagnetic chain as a ring with a missing bond provides an intuitive picture. The cyclic
symmetry of a ring implies a shift quantum number $q =(r-\frac{N}{2}){2 \pi \over N}$, with $r =1,2,\ldots,N$ and
$\hat{T} |q\rangle = e^{-iq} |q\rangle$ ($\hat{T}$ is the shift operator). The ground state belongs to $q = 0$, the
triplet of the $L$ band to $q = \pi$, and the triplets of the $E$ band to $q = \pm\frac{1}{3}\pi$, $\pm\frac{2}{3}\pi$
for a $N = 6$ ring. In the ring, the states with $q$ and $-q$ are degenerate by symmetry, but in \chem{Cr_6} the
perturbation of the missing bond produces a splitting in first order since $\langle-q|\mathcal{V}|+q\rangle \neq 0$.
Accordingly, the wavefunctions are no longer eigenstates of $q$ (running waves), they are given by the symmetric and
antisymmetric mixtures
\begin{equation}
 |q\rangle_\pm = |+q\rangle \pm |-q\rangle,
\end{equation}
as depicted in the inset of fig.~\ref{fig:IN5}. These wavefunctions can be considered as the cosine and sine components
built from $|-q\rangle$ and $|+q\rangle$, in analogy to $e^{-iq} \pm e^{iq}$, and hence represent standing waves. In
\chem{Cr_6} the splitting between them is given by $\Delta_\pm =
2|\langle-\frac{\pi}{3}|\mathcal{V}|\frac{\pi}{3}\rangle| =
2|\langle-\frac{2\pi}{3}|\mathcal{V}|\frac{2\pi}{3}\rangle|$, and the $E$-band triplets organize into two groups
separated by $\Delta_\pm$, see fig.~\ref{fig:SWvsSSW}a (the small splitting of the two states within one group, which
increases with $S$, is a higher-order effect). Obviously, the observation of peaks IV and V in \chem{Cr_6},
fig.~\ref{fig:IN5}, directly reflects the splitting $\Delta_\pm$ due to the formation of the $|q\rangle_\pm$ states,
i.e., the observation of standing spin waves in a finite antiferromagnetic chain.

\begin{figure}
 \onefigure[scale=1]{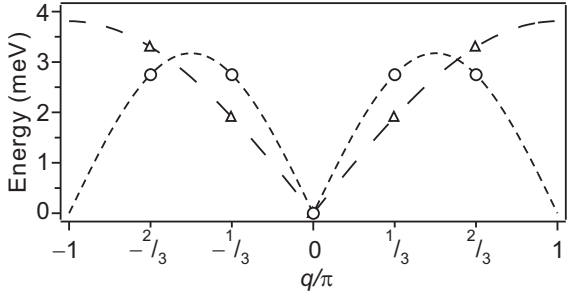}
\caption{Spin-wave dispersions vs $q$ obtained from linear SWT for chains with periodic (dotted line) and open (dashed
line) boundary conditions. Symbols indicate the predicted energies for \chem{Cr_6} ($\triangle$) and a $N$=6 ring
($\circ$).}
 \label{fig:SW_calc}
\end{figure}

Another simple rational for the $E$-band splitting is obtained from linear SWT. The dispersion relations for an
infinite antiferromagnetic chain with periodic or open boundary conditions (BC), respectively, are
\cite{Anderson52,McGurn83}
\begin{eqnarray}
 \mbox{periodic BC:} & \hbar\omega_q &=2S_{i}|J_{ring}\sin(q)|, \cr
 \mbox{open BC:} & \hbar\omega_q &=2S_{i}|J\sin(q/2)|.
\end{eqnarray}
They are displayed in fig.~\ref{fig:SW_calc}. In a finite spin cluster, however, only the discrete waves with the $q$'s
restricted to the above values are possible. The spin-wave energies for a $N = 6$ system are indicated in
fig.~\ref{fig:SW_calc} by open symbols. For an antiferromagnetic $N$=6 ring, four almost degenerate spin waves are
predicted, while for an antiferromagnetic $N$=6 chain the spin waves are organized into two groups, exactly as was
just discussed.


As pointed out already in the above, the rotational-band structure implies a classical spin structure in the $L$ and
$E$ bands. This suggests to compare the results of various semi-classical theories (the details of our calculations
will be reported elsewhere). The most widely recognized semi-classical theory in magnetism is SWT. Here, one starts
from the classical ground state (which in the present case is the N\'eel state with all spins oriented up or down) and
accounts for quantum fluctuations by replacing the spin operators by Holstein-Primakoff or Dyson-Maleev bosons
\cite{Ivanov04}. This yields a boson Hamiltonian $\mathcal{H}_B$ in real space, which is the appropriate starting
point for finite spin systems as pointed out in Ref.~\cite{Cepas05}.

\begin{figure}
 \onefigure[scale=1]{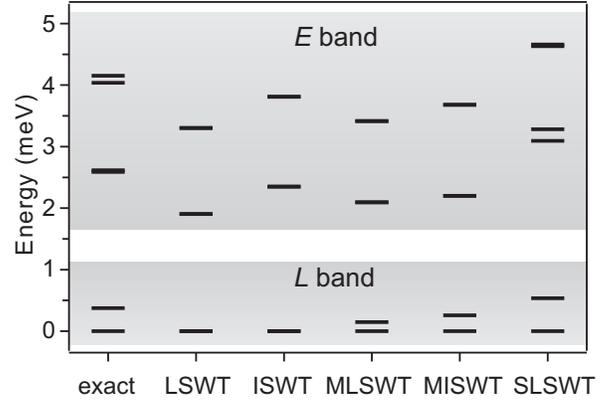}
\caption{Calculated levels for an antiferromagnetic $N = 6$ spin-3/2 chain comparing the exact result and the results
of the SWT calculations for $J = -1.27\un{meV}$ (LSWT = linear SWT, ISWT = interacting SWT, MLSWT = modified-linear
SWT, MISWT = modified-interacting SWT, SLSWT = spin-level SWT). The shaded areas highlight the two lowest $L$-band
states and the $E$-band triplets.} \label{fig:QM_SW}
\end{figure}

In linear SWT, the boson Hamiltonian $\mathcal{H}_B$ is truncated to terms quadratic in the boson operators. The
spin-wave energies are then obtained from a Bogoliubov diagonalization (in this work, the Bogoliubov diagonalization
was performed numerically for all SWT calculations). Interacting SWT additionally includes the effects of 4th-order
boson terms \cite{Ivanov04}.

A serious drawback of these kind of theories is that they start from the assumption of an ordered, symmetry-broken
ground state, which for finite clusters is obviously incorrect. As a result they are unable to reproduce the
singlet-triplet gap, which is characteristic for finite spin systems. These theories hence miss the experimentally
observed peak I. Modified SWTs have become popular to correct for this deficiency \cite{Taka87,Cepas05}. Here, the
rotational invariance is restored by imposing the conditions $\langle S^z_i \rangle = 0$ of zero on-site magnetization
via Lagrange multipliers. Modified-linear SWT goes up to quadratic boson terms, and modified-interacting SWT up to
4th-order boson terms.

Finally, a further approach is suggested. For antiferromagnetic rings (and similarly for other antiferromagnetic
clusters such as Fe$_{30}$) the rotational-band structure implies approximating the wavefunctions of the $L$ and $E$
bands by the spin levels $|\alpha S_A S_B S M\rangle$, where $S_A$ and $S_B$ are the total spins on each of the two
antiferromagnetic sublattices $A$ and $B$ ($\alpha$ denotes intermediate spin quantum numbers)
\cite{OW_SPINDYN,Bernu92,Schnack01}. The $L$ band is obtained for $S_A = S_B = S_i N/2$, while the space of the $E$
band is spanned by the spin levels with $S_A = S_i N/2$, $S_B = S_i N/2 - 1$, and vice versa \cite{OW_SPINDYN}. This
suggests to diagonalize $\mathcal{H}$ in the space of the spin levels $|\alpha S_A S_B S M\rangle$ with the above
values for $S_A$, $S_B$. This approach is called spin-level SWT.

The results obtained for \chem{Cr_6} with all these techniques are shown in fig.~\ref{fig:QM_SW} and compared to the
exact values obtained numerically by exact diagonalization. Some general trends can be observed (which have been
confirmed by studies on more spin clusters). First the $E$ band is discussed. For infinite spin systems the experience
is that linear SWT underestimates energies, while interacting SWT often yields very good results \cite{Ivanov04}. For
\chem{Cr_6} a similar trend is observed. The linear SWT results are well below the exact energies, while interacting
SWT comes rather close, within 8\% accuracy for \chem{Cr_6}. The results of the modified SWTs show a similar trend,
i.e. modified-interacting SWT yields a better agreement with the exact result than modified-linear SWT, although not
quite as good as interacting SWT. In contrast to these (bosonic) SWTs, spin-level SWT overestimates the spin-wave
energies, by about 30\% for \chem{Cr_6}. As mentioned before, the modified SWTs yield a finite singlet-triplet gap
$\Delta_0$ as does the spin-level SWT. $\Delta_0$ from modified-interacting SWT turns out 30\% smaller than the exact
result, while the spin-level SWT calculates it 50\% too large (the performance of modified-linear SWT is even worse).

Hence, we observe that the bosonic SWTs underestimate the energies while the spin-level SWT overestimates them. The
interacting SWT yields the best results for the spin-wave energies, but is not able to reproduce a non-zero
singlet-triplet gap (by construction). The modified-interacting and spin-level SWTs both yield a non-zero value for the
singlet-triplet gap $\Delta_0$, but the overall accuracy is modest, with no clear advantage for one or the other. A
conceptional advantage of spin-level SWT though is that it is rotationally invariant, and that it can be naturally
extended to systems with, e.g., three sublattices such as Fe$_{30}$, where the non-collinear spin structures require
special attention in the interacting SWTs.


In conclusion, we have observed the standing spin-wave excitations in the molecular antiferromagnetic finite-chain
compound \chem{Cr_6} experimentally. We systematically investigated the capabilities of various SWTs in describing the
elementary excitations of finite Heisenberg spin clusters. In our opinion, such investigations are of great value with
regard to molecular nanomagnets where exact results are not accessible. The extendability of the rotational-band
concept to spin systems such as finite antiferromagnetic chains underpins a certain generality.

\acknowledgments

We thank O. C\'epas and T. Ziman for many useful discussions. Financial support by EC-RTN-QUEMOLNA, contract n$^\circ$
MRTN-CT-2003-504880, and the Swiss National Science Foundation is acknowledged.


\begin{thebibliography}{0}

\bibitem{Mn12_Fe8}
 \Name{Gatteschi D. \and Sessoli R.}
 \REVIEW{Angew. Chem. Int. Ed.}{42}{2003}{268}.

\bibitem{Taft94}
 \Name{Taft K. L., Delfs C. D., Papaefthymiou G. C., Foner S., Gatteschi D. \and Lippard S. J.}
 \REVIEW{J. Am. Chem. Soc.}{116}{1994}{823}.

\bibitem{Gat94}
 \Name{Gatteschi D., Caneschi A., Pardi L. \and Sessoli R.}
 \REVIEW{Science}{265}{1994}{1054}.

\bibitem{OW_CCR}
 \Name{Waldmann O.}
 \REVIEW{Coordin. Chem. Rev.}{249}{2005}{2550}.

\bibitem{Mueller01}
 \Name{M\"uller A., Luban M., Schr\"oder C., Modler R., K\"ogerler P., Axenovich M., Schnack J., Canfield P. C.,
Budko S. \and Harrison N.}
 \REVIEW{ChemPhysChem.}{2}{2001}{517}.

\bibitem{Schnack01}
 \Name{Schnack J., Luban M. \and Modler R.}
 \REVIEW{Europhys. Lett.}{56}{2001}{863}.

\bibitem{Anderson52}
 \Name{Anderson P. W.}
 \REVIEW{Phys. Rev.}{86}{1952}{694}.

\bibitem{Bernu92}
 \Name{Bernu B. \etal}
 \REVIEW{Phys. Rev. Lett.}{69}{1992}{2590};
 \Name{Lecheminant P. \etal}
 \REVIEW{Phys. Rev. B}{52}{1995}{6647}.


\bibitem{Schnack00}
 \Name{Schnack J. \and Luban M.}
 \REVIEW{Phys. Rev B}{63}{2000}{014418}.

\bibitem{OW_SPINDYN}
 \Name{Waldmann O.}
 \REVIEW{Phys. Rev. B}{65}{2002}{024424}.

\bibitem{OW_Cr8}
 \Name{Waldmann O., Guidi T., Carretta S., Mondelli  C. \and Dearden A. L.}
 \REVIEW{Phys. Rev. Lett.}{91}{2003}{237202}.

\bibitem{Garlea06}
 \Name{Garlea V. O., Nagler S. E., Zarestky J. L., Stassis C., Vaknin D., K\"ogerler P., McMorrow D. F.,
Niedermayer C., Tennant D. A., Lake B., Qiu Y., Exler M., Schnack J. \and Luban M.}
 \REVIEW{Phys. Rev. B}{73}{2006}{024414 and references therein}.
 \Name{Waldmann O.}
 \REVIEW{Phys. Rev. B}{75}{2007}{012415}.

\bibitem{Ivanov04}
 \Name{Ivanov N. B. \and Sen D.}
 \REVIEW{Lecture Notes in Physics}{645}{2004}{195}.

\bibitem{Yama02}
 \Name{Yamamoto S. \and Nakanishi T.}
 \REVIEW{Phys. Rev. Lett.}{89}{2002}{157603}.

\bibitem{Chab04}
 \Name{Chaboussant G., Sieber A., Ochsenbein  S., G\"udel  H.-U., Murrie  M., Honecker A., Fukushima N. \and
Normand B.}
 \REVIEW{Phys. Rev. B}{70}{2004}{104422}.

\bibitem{Cepas05}
 \Name{C\'epas O. \and Ziman T.}
 \REVIEW{Prog. Theo. Phys. Suppl.}{159}{2005}{280}.

\bibitem{chains}
 \Name{Hagiwara M. , Katsumata K., Affleck I., Halperin B. I. \and Renard J. P.}
 \REVIEW{Phys. Rev. Lett.}{65}{1990}{3181};
 \Name{Ditusa J. F., Cheong S. W., Park  J. H., Aeppli G., Broholm C. \and Chen C. T.}
 \REVIEW{Phys. Rev. Lett.}{73}{1994}{1857};
 \Name{Fujiwara N., Yaskuoka  H., Isobe M. \and Ueda Y.}
 \REVIEW{Phys. Rev. B}{58}{1998}{11134};
 \Name{Lou J., Qin S., Ng T.-K., Su Z. \and Affleck I.}
 \REVIEW{Phys. Rev. B}{62}{2000}{3786};
 \Name{Bogani L., Caneschi A., Fedi M., Gatteschi D., Massi M., Novak M. A., Pini G. M., Rettori A., Sessoli R. \and Vindigini A.}
 \REVIEW{Phys. Rev. Lett.}{92}{2004}{207204};
 \Name{Parkinson J. B.}
 \REVIEW{J. Phys.: Condens. Matter}{16}{2004}{S5233};
 \Name{Furrer A. \and G\"udel  H.-U.}
 \REVIEW{Eur. Phys. J. B}{16}{2000}{81}.


\bibitem{Larsen03}
 \Name{Larsen F. K., Overgaard J., Parsons S., Rentschler E., Smith A. A., Timco G. A. \and Winpenny R. E. P.}
 \REVIEW{Angew. Chem. Int. Ed.}{42}{2003}{5978}.

\bibitem{Cr6pair}
 \Page{The anions in \chem{Cr_6} are arranged in pairs separated by six \chem{[^nPr_2NH_2]^+} cations; \chem{^nPr} stands for
n-propyl.}

\bibitem{Cador04}
 \Name{Cador O., Gatteschi D., Sessoli R., Larsen F. K., Overgaard  J., Barra A.-L., Teat S. J., Timco G. A. \and Winpenny R.
E. P.}
 \REVIEW{Angew. Chem. Int. Ed.}{43}{2004}{5196}.

\bibitem{phonons}
 \Page{The non-magnetic origin was shown by correcting the data by a Bose-population factor for phonons and comparing different
temperatures.}

\bibitem{OW_INS}
 \Name{Waldmann O.}
 \REVIEW{Phys. Rev. B}{68}{2003}{174406}.

\bibitem{smallpks}
 \Page{The three very weak peaks appearing in the calculated spectrum at 2.2, 3.0 and 3.5$\un{meV}$ stem from
transitions too weak to be observed experimentally, and hence are not labeled.}

\bibitem{Cr6params}
 \Page{$J_{12} = J_{56} = -1.1\un{meV}$, $J_{23} = J_{34} = J_{45} = -1.4\un{meV}$. The Cr ions Cr1 and Cr6 have a
 different coordination sphere than the Cr ions Cr2 $-$ Cr5. The coordination spheres affect the nearest-neighbor
 exchange interactions, hence $J_{12}$ and $J_{56}$ are expected to be slightly different from $J_{23}$, $J_{34}$, and
 $J_{45}$ (more details will be reported elsewhere).}

\bibitem{McGurn83}
 \Name{McGurn A. R. \and Thorpe M. F.}
 \REVIEW{J. Phys. C: Solid State Phys.}{16}{1983}{1255}.

\bibitem{Taka87}
 \Name{Takahashi M.}
 \REVIEW{Phys. Rev. Lett}{58}{1987}{168};
 \Name{Hirsch J. E. \and Tang S.}
 \REVIEW{Phys. Rev. B}{40}{1989}{4769}.

\end{thebibliography}
\end{document}